\documentclass[journal]{IEEEtran}

\usepackage{etex}
\usepackage[caption=false,font=footnotesize]{subfig}
\usepackage{amsmath}
\usepackage{float}
\usepackage{graphicx}
\usepackage{cite}
\usepackage{fancyhdr}
\usepackage{eso-pic}

\makeatletter

\let\old@ps@headings\ps@headings
\let\old@ps@IEEEtitlepagestyle\ps@IEEEtitlepagestyle
\def\confheader#1{%
    \def\ps@headings{%
        \old@ps@headings%
        \def\@oddfoot{\strut\hfill#1\hfill\strut}%
        \def\@evenfoot{\strut\hfill#1\hfill\strut}%
        \def\@oddhead{\strut\hfill Accepted for publication: IEEE/OSA Journal of Lightwave Technology \hfill\strut}%
        \def\@evenhead{\strut\hfill Accepted for publication: IEEE/OSA Journal of Lightwave Technology \hfill\strut}%
    }%
    \def\ps@IEEEtitlepagestyle{%
        \old@ps@IEEEtitlepagestyle%
        \def\@oddhead{\strut\hfill Accepted for publication: IEEE/OSA Journal of Lightwave Technology \hfill\strut}%
        \def\@evenhead{\strut\hfill Accepted for publication: IEEE/OSA Journal of Lightwave Technology \hfill\strut}%
				\def\@oddfoot{\strut\hfill#1\hfill\strut}%
        \def\@evenfoot{\strut\hfill#1\hfill\strut}%
    }%
    \ps@headings%
}
\makeatother

\confheader{%
        \parbox{15cm}{\footnotesize \begin{center} Copyright (c) 2018 IEEE. Personal use of this material is permitted.  However, permission to use this material for any other purposes must be obtained from the IEEE by sending a request to pubs-permissions@ieee.org. \end{center}}
}

\begin{document}

\title{Interplay of Probabilistic Shaping and the Blind Phase Search Algorithm}

\author{Darli A. A. Mello, ~Fabio A. Barbosa and Jacklyn D. Reis
\thanks{D. A. A. Mello and F. A. Barbosa are with the School of Electrical and Computer Engineering, University of Campinas (Unicamp), Campinas, Brazil.}
\thanks{J. D. Reis is with Idea! Electronic Systems, Campinas, Brazil.}
\thanks{At Unicamp, this work was supported by FAPESP grants 2015/24341-7 and 2015/24517-8. J. Reis was supported by CNPq grant 311871/2016-0. We would like to thank Omar Domingues for the constrained capacity calculations. 

Part of this work appears in \cite{Barbosa2018}.}}

\maketitle

\begin{abstract}
Probabilistic shaping (PS) is a promising technique to approach the Shannon limit using typical constellation geometries. However, the impact of PS on the chain of signal processing algorithms of a coherent receiver still needs further investigation. In this work we study the interplay of PS and phase recovery using the blind phase search (BPS) algorithm, which is widely used in optical communications systems. We first investigate a  supervised phase search (SPS) algorithm as a theoretical upper bound on the BPS performance, assuming perfect decisions.  It is shown that PS influences the SPS algorithm, but its impact can be alleviated by moderate noise rejection window sizes. On the other hand, BPS is affected by PS even for long windows because of correlated erroneous decisions in the phase recovery scheme. The simulation results also show that the capacity-maximizing shaping is near to the BPS worst-case situation for square-QAM constellations, causing potential implementation penalties. 
\end{abstract}


\IEEEpeerreviewmaketitle

\section{Introduction}

Probabilistic shaping (PS) is a digital transmission technique by which constellation symbols are transmitted with different a-priori probabilities. In general, symbols with larger amplitudes are transmitted with lower probabilities. PS maximizes the mutual information (MI) achieved by the transmission scheme for a given signal constellation and signal to noise ratio (SNR) and allows, in certain conditions, to approach the Shannon limit. Although PS has been known for decades \cite{Kschischang93,Wachsmann99}, its application on practical systems is still in its infancy. Significant implementation advances have been recently proposed by B\"ocherer et al. in \cite{Bocherer2015}. 

In optical systems, the interest in PS has gained significant momentum. To our knowledge, PS has been first addressed in the context of optical communications by Beygi et al. in \cite{Beygi2014}, where a  rate-adaptive coded modulation scheme with probabilistic signal shaping has been proposed. The impact of rate-adaptive coded modulation with PS on optical networking has been quantified by Mello et al. in \cite{Mello14}. Yankov et al. have investigated in \cite{Yankov2014} an implementation of PS for turbo codes. The combination of PS with low-density parity-check  codes (LDPC) for optical communications has been shown by Fehenberger et al. in \cite{Fehenberger2015}. The first experimental demonstration of PS for optical communications has been accomplished by Buchali et al. in \cite{Buchali2015}, for a 64-QAM modulated signal. Since then, PS has been applied to different contexts, ranging from transoceanic applications \cite{Domingues2017,Ghazisaeidi2016} to unrepeatered optical transmission \cite{Renner2017}. PS has already been demonstrated in a large set of experiments, but fully supervised equalization and phase recovery, with controlled  conditions, are largely used. One of the first works to relate phase recovery and PS in more practical scenarios has been recently presented by Pilori et al. in \cite{Pilori2018}. Supervised and partially-supervised pilot-aided phase recovery were investigated. Supervision using 2\% pilot overhead is applied to the phase unwrapper to mitigate cycle slips. The pilot-aided scheme achieved equivalent performance as the supervised scheme at linear propagation regimes, but exhibited some penalty in the presence of nonlinear interference. However, the performance of phase recovery algorithms was assessed from an end-to-end perspective and in particular configurations. 

\begin{figure*}[!htb]
\centering
\includegraphics[scale = 1]{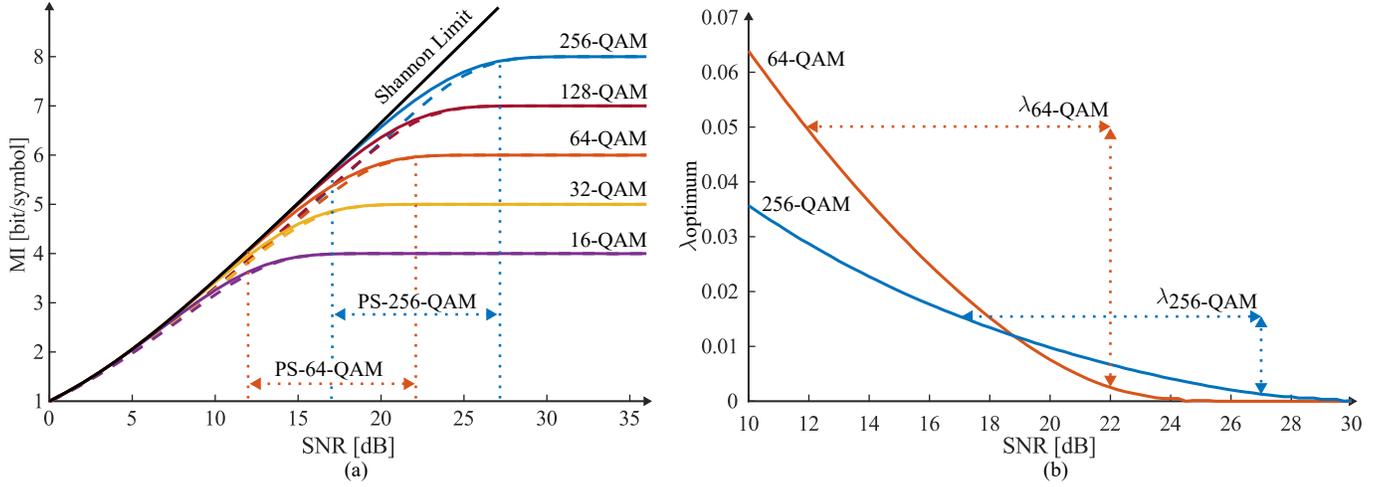}
\caption{(a) MI for typical M-QAM formats. Dashed lines: uniform constellations. Solid lines: probabilistically shaped constellations. The dotted lines indicate the interval of interest for PS-64-QAM and PS-256-QAM. (b) Optimum values of $\lambda$ for PS-64-QAM and PS-256-QAM. The dotted lines indicate the range of $\lambda$s that corresponds to the interval of interest shown in Fig. \ref{figure:mutualinformation_lambdaopt}(a). Note that $\lambda = 0$ corresponds to a uniform constellation.} \label{figure:mutualinformation_lambdaopt}
\end{figure*}

In \cite{Barbosa2018}, we have shown that PS can affect the performance of the blind phase search (BPS) algorithm, which is widely used in optical communications systems. The performance of the algorithm was evaluated by simulations. In this paper, we extend the results of \cite{Barbosa2018}, and provide a detailed analysis on the interplay of PS and BPS. Supervised phase search (SPS), a phase recovery algorithm with the same architecture of BPS, but with perfect decisions, is investigated by analytical derivations and Monte Carlo simulations. This configuration is used to derive an upper bound on the BPS performance. BPS is only studied by simulations, as the analytical modeling becomes overly complex because of the decision process. As in \cite{Barbosa2018}, the investigated phase recovery algorithms are first assessed by the mean square error (MSE) of a constant phase shift estimated over a given observation window, in a way to allow an analytical treatment of the problem. Subsequently, we investigate the impact of phase recovery on the mutual information (MI) of a channel with Wiener phase noise. Here, eventual cycle slips are circumvented by a supervised phase unwrapper.

The remainder of this paper is divided as follows. Section \ref{section:system} details the system model, including the PS technique and the BPS and SPS algorithms. Section \ref{simulation} presents the simulation setup and results. Lastly, Section \ref{section:conclusion} concludes the paper.

\section{System model} \label{section:system}
\subsection{Probabilistic shaping (PS)}

Probabilistic shaping is usually implemented by applying the  Maxwell-Bolzmann  distribution to the a-priori probabilities $P_m$ of symbols $s_m$ of the transmitted constellation \cite{Kschischang93}:
\begin{equation}
\displaystyle P_m = \frac{e^{-\lambda |s_m|^2}}{\sum_{k=1}^M e^{-\lambda |s_k|^2}} \label{equation:MB}
\end{equation}
where $\lambda$ is the shaping parameter and $M$ is the constellation size. The choice of $\lambda$ must be made carefully, as the optimum value varies according to the signal power, modulation format and SNR. Fig. \ref{figure:mutualinformation_lambdaopt}(a) shows the MI for typical modulation formats with uniform (dashed line) and shaped (solid line) constellations. For the sake of clarity, we focus in this paper on the PS-64-QAM and PS-256-QAM formats, but the analysis can also be easily extended to other schemes. Fig. \ref{figure:mutualinformation_lambdaopt}(a) helps to understand the range of SNRs for which shaping should be applied for a specific modulation format. For PS-64-QAM, for example, PS should not be applied for SNRs higher than 22 dB, as uniform and shaped constellations achieve the same MI. On the other hand, PS should not be deployed with an SNR below 12 dB, as PS-32-QAM achieves equivalent performance causing a potentially lower implementation penalty. An analogous analysis can be carried out for the PS-256-QAM format, for which the SNR interval of interest ranges from 17 dB to 27 dB. Fig. \ref{figure:mutualinformation_lambdaopt}(b) shows the optimum $\lambda$ parameter for the PS-64-QAM and PS-256-QAM formats, with in-phase and quadrature components having amplitudes\footnote{Note that the choice of $\lambda$ depends on the signal power.} $\pm (2i+1), i=0,1,...,\sqrt{\textrm{M}}/2-1$ . The figure allows to infer the range of $\lambda$ for which PS should be implemented, namely, from 0 to 0.05 for PS-64-QAM and from 0 to 0.015 for \mbox{PS-256-QAM}. Note that, $\lambda=0$ corresponds to a uniform constellation.

\subsection{Supervised and blind phase search algorithms (SPS and BPS) }

Let the $i^{th}$ constellation symbol $s_i$ be transmitted over a complex additive white Gaussian noise (AWGN) channel. The phase noise associated with the transmitter and local oscillator lasers is expressed by a multiplicative factor $e^{j\theta_n}$, so that the received symbol $r_i$ is given by:
\begin{equation}
r_i = s_ie^{j\theta_n}+n'_i
\end{equation}
where the complex Gaussian noise term $n'_i$ has zero mean and variance $2\sigma_n^2$. We define the signal to noise ratio (SNR) as \mbox{$\textrm{SNR} = P_s/2\sigma_n^2$}, where $P_s = E\{ |s_i|^2\}$. Phase recovery algorithms resort to the fact that $\theta_n$ varies slowly over time, in such a way that it is approximately constant over $N$ symbols. In practice, the size of $N$ also depends on the symbol rate and on the linewidth of transmitter and local oscillator lasers.

The BPS algorithm estimates the phase noise rotation $\theta_n$ as the angle that minimizes the sum of squared distances between $N$ adjacent symbols $s_i$, rotated by a test phase $\theta_r$, and their respective estimates $\hat{s}_i$. In this section we assume an infinite number of test phases and do not delve into resolution issues.   In mathematical terms, estimate $\hat{\theta}_n$ is obtained as:
\begin{equation}
\displaystyle \hat{\theta}_{n} = \min_{\theta_r}J(\theta_r)
\end{equation}
where the cost function $J(\theta_r)$ is given by:
\begin{eqnarray}
J(\theta_r) &=& \sum_{i=1}^N | e^{-j\theta_r}(s_ie^{j\theta_n}+n'_i) - \hat{s}_i|^2 \\
&=& \sum_{i=1}^N | s_ie^{j(\theta_n-\theta_r)} - \hat{s}_i+n_i|^2 \label{equation:optimization}
\end{eqnarray}
\indent Term $n_i$ is a rotated Gaussian process of the same mean and variance of $n'_i$. It is also possible to write $J(\theta_r)$ as a function of the symbol error $e_i=s_i-\hat{s_i}$:
\begin{equation}
J(\theta_r) = \sum_{i=1}^N | s_ie^{j(\theta_n-\theta_r)} - s_i+e_i+n_i|^2 \label{equation:J}
\end{equation}
\indent In order to provide an analytical insight to the problem, we first investigate an ideal algorithm called SPS, which follows the same steps of BPS, except for the fact that the algorithm is not affected by erroneous decisions. In practical implementations, SPS can be deployed in bursts to periodically refresh BPS. In SPS $e_i=0$  and the analytical modeling is simplified. It can be shown that the MSE of SPS in the estimation of $\theta_n$ can be approximated by (see Appendix A for the complete derivation):
\label{section:analytical}
\begin{equation}
 \displaystyle \textrm{MSE}_{\textrm{SPS}}(N) = E\{(\theta_n - \hat{\theta}_n )^2\} \approx    E\left \{\left[\frac{\sum_{i=1}^N (n_i^{(1)})|s_i|}{\sum_{i=1}^N |s_i|^2} \right]^2\right \} \label{Equation:MSE} 
\end{equation}
where $n_i^{(1)}$ is the noise component in the direction of the subtraction of $s_i$ and its rotated version $s_ie^{j(\theta_n-\theta_r)}$. 

The computation of (\ref{Equation:MSE}) is not trivial for intermediate values of $N$, but the extreme cases offer interesting insights. Setting $N=1$ gives:
\begin{equation}
    \textrm{MSE}_{\textrm{SPS}}(1) \approx  E\left \{\left[\frac{n_i^{(1)}}{ |s_i|} \right]^2\right \}  \\
     =  \sigma_n^2 \sum_{m=1}^M \frac{1}{|s_m|^2}P_m \\ \label{equation:nobuffer}
\end{equation}
\indent Clearly, for small windows the SPS performance  depends not only on the SNR, but also on the a-priori probability distribution of transmitted symbols. In communications systems with PS implemented by the Maxwell-Bolzmann distribution, it can be shown, by differentiating (\ref{equation:nobuffer}) with respect to $\lambda$ and setting the result equal to zero, that the MSE is maximized by the following condition:
 \begin{equation}
 \displaystyle \left[ E\{|s_i|^4\}-2E \{|s_i|^2\}^2\right]E\left \{\left|\frac{1}{s_i}\right|^2\right \}+E\{|s_i|^2\}=0 \label{equation:lambdamax}
 \end{equation}
\indent In the derivation of (\ref{equation:lambdamax}), it should be noted that  \mbox{$\sigma_n^2=P_s/(2\textrm{SNR})$}, where $P_s = E\{ |s_i|^2\}$, also depends on $\lambda$. By inspection of (\ref{equation:lambdamax}) one can observe that the SPS performance is affected by several moments of $|s_i|$ and $1/|s_i|$, including the fourth central moment of $|s_i|$. For M-QAM constellations, where $E\{s_i^2\}=0,$ its possible to rewrite  (\ref{equation:lambdamax}) in terms of its Kurtosis, given by $K_s = E\{|s_i|^4\} - 2E^2\{|s_i|^2\} - |E\{s_i^2\}|^2$. Thus, the MSE is maximized when:
 \begin{equation}
 K_s = -\frac{E\{|s_i|^2\}} {E\left \{\left|1/{s_i}\right|^2\right \}}
 \end{equation}
\indent On the other hand, supposing a large value of $N$, called here $N_L$, the Law of Large Numbers can be invoked to assume that, in the observation window, $N_LP_m$  symbols of type $s_m$ occur. Thus, MSE$_{\textrm{SPS}}$($N_L$) becomes:       
\begin{align}
\displaystyle  \textrm{MSE}_{\textrm{SPS}}(N_L)  &\approx \frac{\sigma_n^2 N_L \sum_{m=1}^M |s_m|^2P_m}{N_L^2(\sum_{m=1}^M |s_m|^2P_m)^2} \\
& =  \frac{\sigma_n^2}{N_LP_s} = \frac{1}{2N_L}\textrm{SNR}^{-1} \label{equation:SPS}
\end{align}
\indent Thus, for large  window sizes, the SPS performance depends on the SNR, but is weakly affected by the transmitted constellation. This can be explained by the sums of $N_L$ independent and identically distributed random variables in (\ref{Equation:MSE}), allowing us to invoke the Central Limit Theorem.

In BPS, $e_i \neq 0$, and the analytical modeling becomes challenging because $e_i$ depends on  $n_i$, $s_i$ and  $\theta_r$. Therefore, the analysis of BPS is carried out by simulation.

\section{Simulation setup and results} \label{simulation}
\subsection{MSE performance}

We assume that shaping changes the a-priori probability of transmitted symbols, but keeps their location in the complex plane in $\pm (2i+1), i=0,1,...,\sqrt{\textrm{M}}/2-1$. This assumption has a direct influence in the choice of $\lambda$, as it depends on the constellation amplitudes (although  $P_m$ depends only on the SNR). In practice, the absolute values of signal and additive noise powers are meaningless for the phase recovery algorithm, as only the SNR dictates the transceiver performance. Monte Carlo simulations were carried out considering $2^{19}$ symbols. An arbitrary constant rotation of $\pi/6$ rad was applied to the symbols, to represent a constant phase noise in a given window. Thus, the larger the window size, the better the performance of the estimation algorithm. In practical applications, the optimum window size depends on the system operating conditions, such as the optical signal to noise ratio (OSNR), laser linewidth, and symbol rate. In this section, the size of the window was varied to simulate these different operating conditions without entering into system issues. As the phase deviation is kept constant throughout the simulation in the first quadrant, there is no need to implement a phase unwrapper after BPS. AWGN was added to the generated signals to guarantee a constant SNR, independently of the amount of shaping applied to the constellation.  To circumvent resolution issues, 900 test phases are used in the SPS and BPS algorithms. 

\begin{figure*}
\centering
\includegraphics[scale = 1]{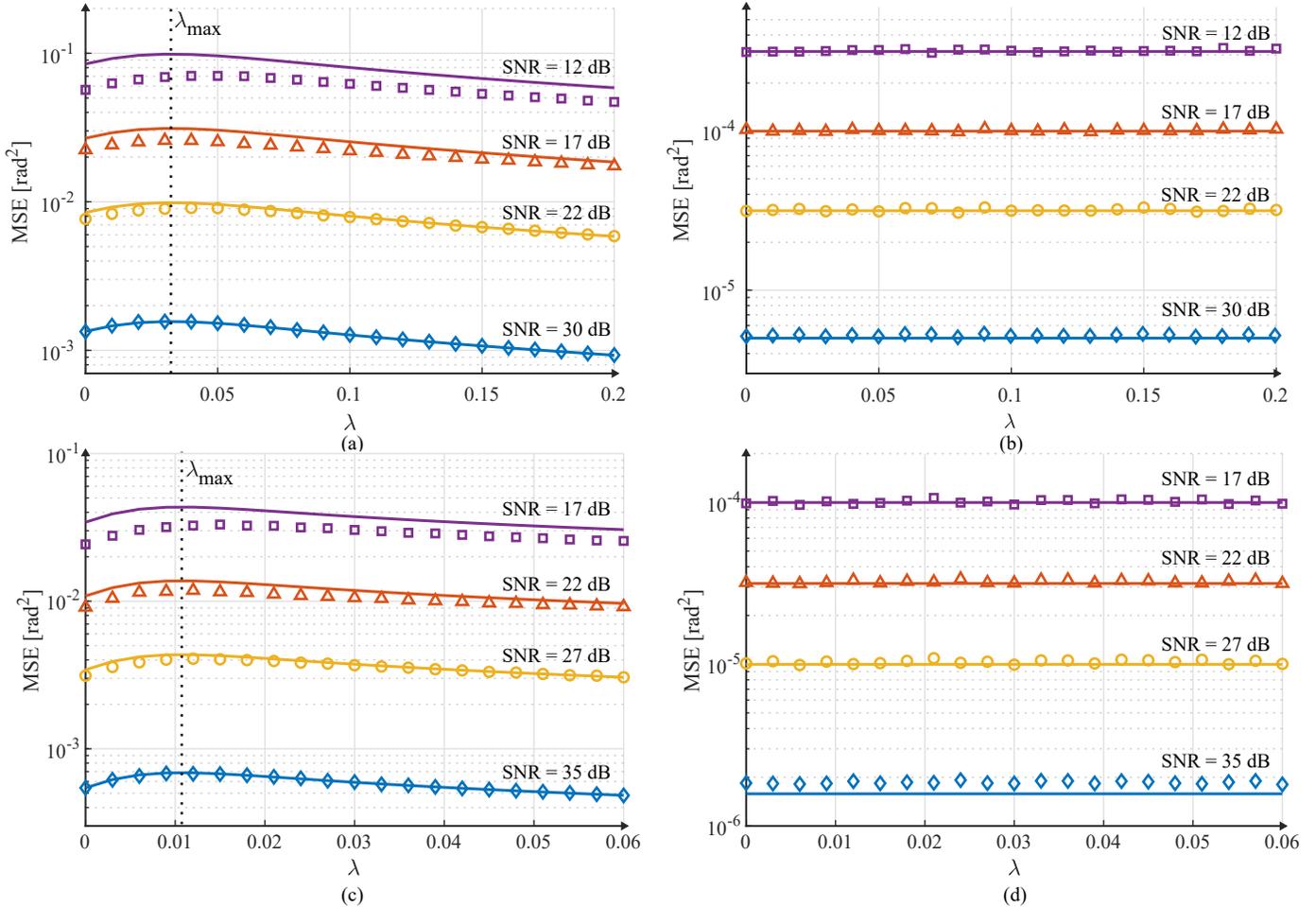}
\caption{MSE for SPS with PS-64-QAM and (a) $N$ = 1 and (b) $N$ = 100. MSE for PS-256-QAM with (c) $N$ = 1 and (d) $N$ = 100. The solid lines indicate analytical results obtained by MSE$_{\textrm{SPS}}$(1) and MSE$_{\textrm{SPS}}$($N_L=100$), while the symbols were generated by Monte Carlo simulations. The transmission channel includes AWGN and a constant phase shift of $\pi/6$. The curves indicate that moderate noise rejection windows are sufficient to make the SPS performance independent on PS. The vertical dotted lines in figures (a) and (c) indicate $\lambda_{\textrm{max}}$, calculated analytically by (\ref{equation:lambdamax}). Note that $\lambda=0$ corresponds to a uniform constellation.} \label{figure:supervised}
\end{figure*}

\begin{figure*}
\centering
\includegraphics[scale = 1]{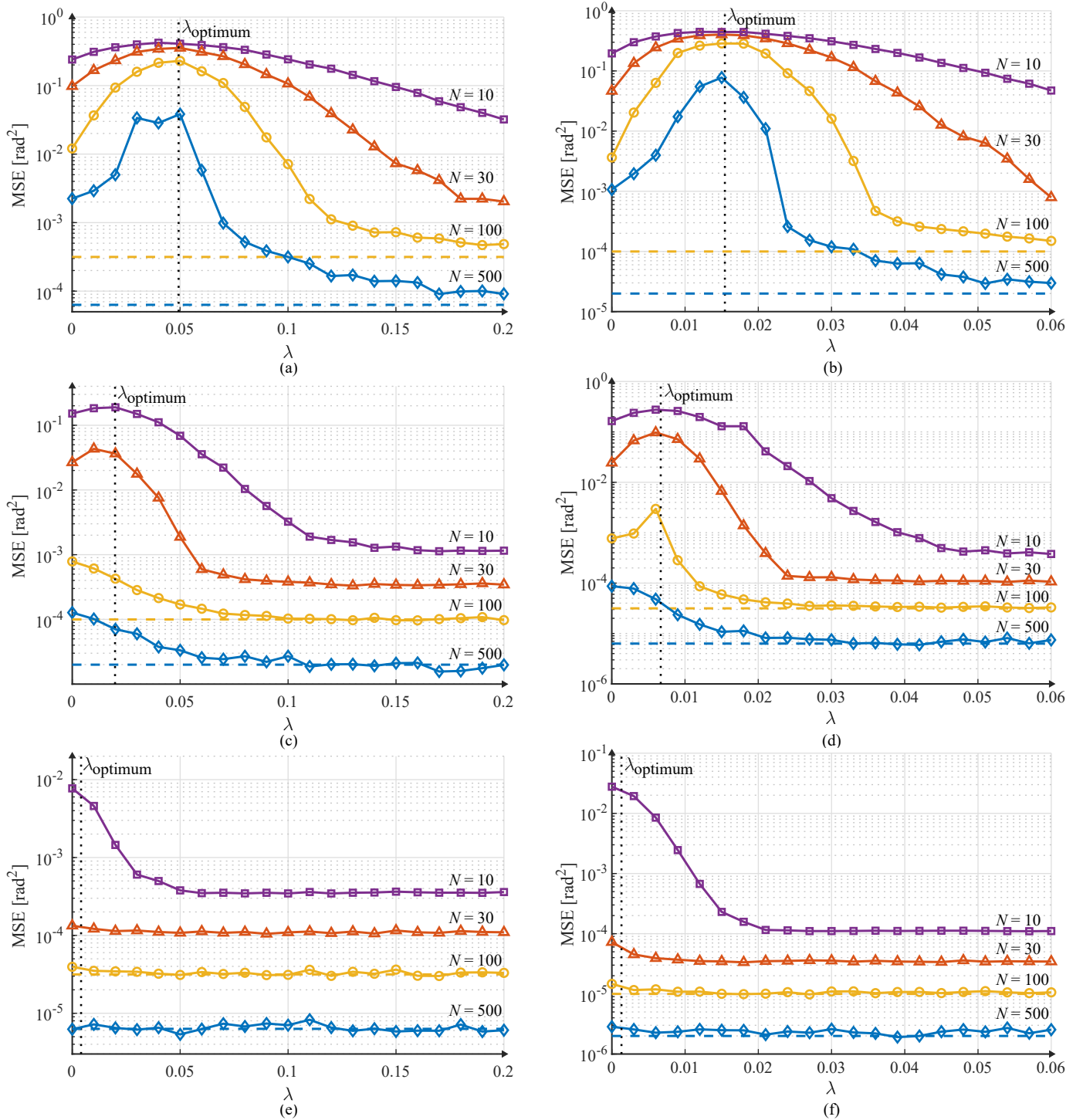} 
\caption{MSE for BPS with $N$ = 10, 30, 100, and 500, evaluated with PS-64-QAM at (a) SNR = 12 dB, and (c) SNR = 17 dB, and (e) SNR = 22 dB; and evaluated with PS-256-QAM at (b) SNR = 17 dB, (d) SNR = 22 dB, and (f) SNR = 27 dB. The transmission channel includes AWGN and a constant phase shift of $\pi/6$. The dotted lines indicate $\lambda_{\textrm{optimum}}$ for the corresponding configuration.  The dashed lines indicate the SPS predictions for large $N$, for $N=100$ and $N=500$, given by (\ref{equation:SPS}). Note that $\lambda=0$ corresponds to a uniform constellation.} \label{figure:mseunsupervised}
\end{figure*}

\begin{figure*}
\centering
\includegraphics[scale = 1]{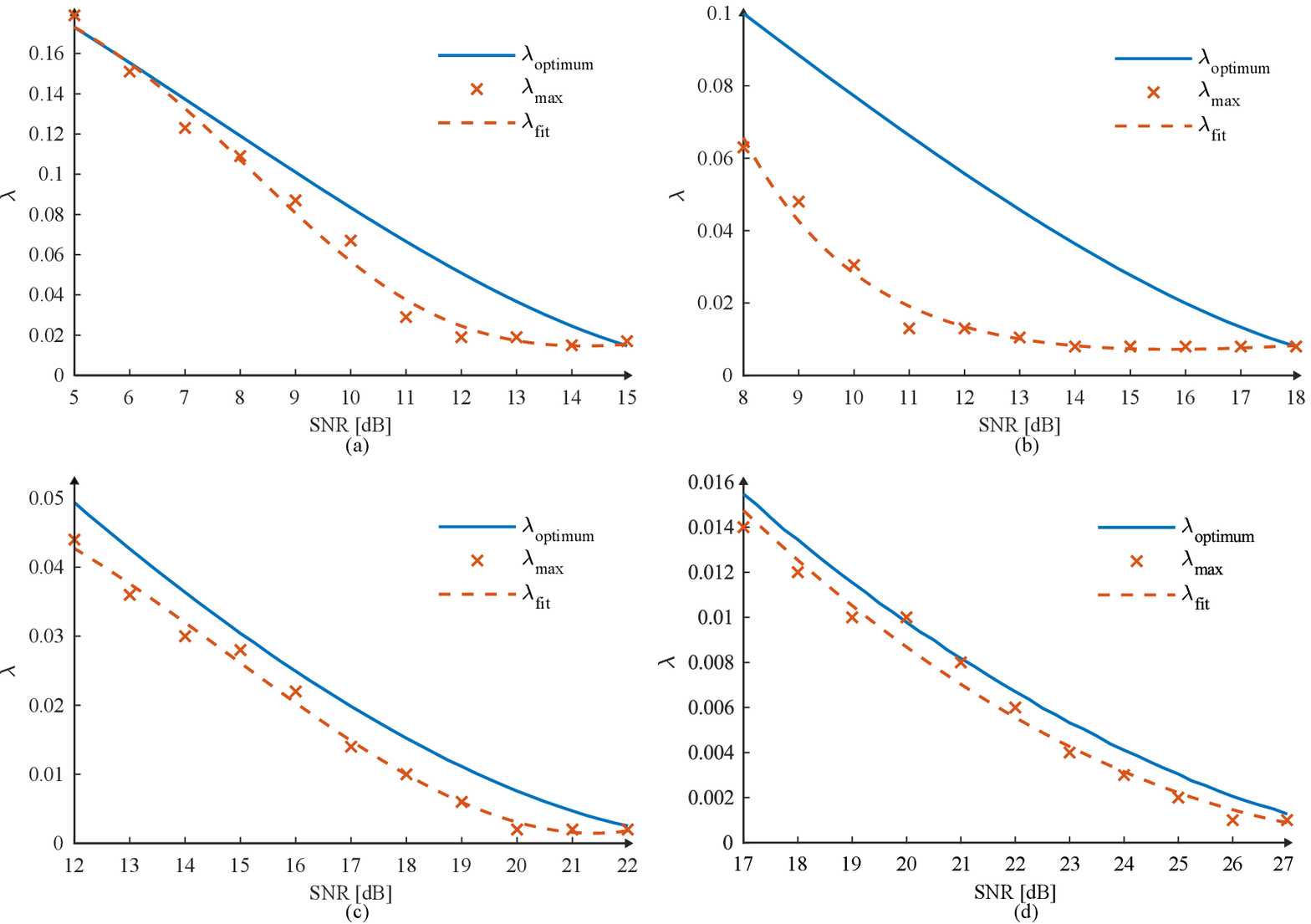} 
\caption{Theoretical $\lambda_{\textrm{optimum}}$ for $N$ = 10 (solid lines) and simulated $\lambda_{\textrm{max}}$ (markers) for  (a) PS-16-QAM, (b) PS-32-QAM, (c) PS-64-QAM, and (d) PS-256-QAM. The transmission channel includes AWGN and a constant phase shift of $\pi/6$. The dashed lines correspond to fitted values of $\lambda_{\textrm{max}}$ ($\lambda_{\textrm{fit}}$).}
\label{figure:lambdamax}
\end{figure*}

Fig. \ref{figure:supervised} shows the MSE of $\theta_n$ for SPS. The solid lines indicate analytical predictions, while the symbols correspond to the results produced by Monte Carlo simulations. The results for SNR = 30 dB and SNR = 35 dB  were included as a high-SNR reference. Figs. \ref{figure:supervised}(a) and \ref{figure:supervised}(b) show the results for the PS-64-QAM format and $N$ = 1 and $N$ = 100, respectively. The analytical approximation for $N$ = 1 exhibits a good agreement with the simulations, with increasing accuracy for higher SNRs. At $N$ = 100 the model accuracy is preserved even at lower SNRs. The same behavior is observed for the PS-256-QAM format in Figs. \ref{figure:supervised}(c) and \ref{figure:supervised}(d). 

As predicted by the analytical model, for \mbox{$N$ = 1} the MSE can increase as a result of shaping compared with the uniform distribution. There are two main processes that explain the shape in Figs. \ref{figure:supervised}(a) and \ref{figure:supervised}(c). To understand them, let us once again assume that in the shaping process the position of the constellation symbols is retained, but its frequency is altered. In the first process, an increasing $\lambda$ reduces the occurrence of large amplitude symbols, impairing the BPS performance. This occurs because phase deviations are more easily detected in large amplitude symbols. In the second process, shaping reduces the signal power and, to maintain the SNR constant, the additive noise power is also downscaled, helping  the estimation process. The dominance of the first process for low $\lambda$ values, and of the second process for high $\lambda$ values, explains the existence of a maximum in the MSE curves. This dependence of the SPS performance on shaping can be easily alleviated by longer noise rejection windows, for which the MSE is practically independent on the modulation format. The figures for $N$ = 1 also show the $\lambda$ parameter value which maximizes the MSE$_{\textrm{SPS}}$ ($\lambda_{\textrm{max}}$) calculated by (\ref{equation:lambdamax}). 

Figs. \ref{figure:mseunsupervised}(a) and \ref{figure:mseunsupervised}(b) show the MSE as a function of $\lambda$, for BPS evaluated with PS-64-QAM at SNR = 12 dB and with PS-256-QAM at SNR = 17 dB, respectively. The horizontal dashed lines show the analytical predictions for SPS with large $N$, obtained by ($\ref{equation:SPS}$), for $N$ = 100 and $N$ = 500. Clearly, BPS is affected by a third process at low SNRs, which is directly influenced by the two processes described for the SPS. It is the generation of decision errors in the estimation of the transmitted symbol. Longer noise rejection windows reduce the MSE, but the filtering gains depend strongly on $\lambda$. For example, for the 64-QAM format without shaping, increasing $N$ from 30 to 100 produces a 10-fold reduction on the MSE. On the other hand, this gain is strongly reduced if the system operates at \mbox{$\lambda$ = 0.05}. A similar trend can be observed for the PS-256-QAM modulation at \mbox{$\lambda$ = 0.015}.  It is interesting to note that, for both PS-64-QAM and PS-256-QAM formats, the maximum MSE is achieved near $\lambda_{\textrm{optimum}}$. Figs. \ref{figure:mseunsupervised}(e) and \ref{figure:mseunsupervised}(f) show the MSE for BPS evaluated with PS-64-QAM at \mbox{SNR = 22 dB} and with \mbox{PS-256-QAM} at SNR = 27 dB, respectively, which are the highest SNR values for which shaping should be applied. Interestingly, for both conditions, in most cases the MSE remains constant or decreases with $\lambda$, indicating that PS can improve the BPS performance. Figs. \ref{figure:mseunsupervised}(c) and \ref{figure:mseunsupervised}(d) are intermediate cases, where the MSE is evaluated with \mbox{PS-64-QAM} at \mbox{SNR = 17 dB} and with PS-256-QAM at \mbox{SNR = 22 dB}. In all observed cases, $\lambda_{\textrm{optimum}}$ approaches the worst-case condition for BPS. This effect was not present in the SPS analysis, for which a moderate noise rejection window was enough to mitigate the impact of PS on the MSE. Therefore, we conjecture that the capacity-maximizing shaping is near to the the worst-case condition for the decision process inside the BPS algorithm. To evaluate this trend, we simulate BPS with a window $N = 10$, and find $\lambda_{\textrm{max}}$ for each SNR. The obtained $\lambda_{\textrm{max}}$ is compared with $\lambda_{\textrm{optimum}}$. Previous analyses in this paper have focused on the \mbox{PS-64-QAM} and \mbox{PS-256-QAM} formats. Here, in order to increase the comprehensiveness of the results, we also  evaluate the PS-32-QAM format -- built by pruning the previously defined \mbox{PS-64-QAM} constellation -- and the \mbox{PS-16-QAM} format, with amplitudes \mbox{$\pm (2i+1), i=0,1$}. The results are shown in Fig. \ref{figure:lambdamax}. It is observed that, for the 16/64/256-QAM constellations, $\lambda_{\textrm{max}}$ is in the vicinity of $\lambda_{\textrm{optimum}}$ in the full range of SNRs, indicating that BPS can cause implementation problems in systems with probabilistic shaping. However, this effect  is not observed for PS-32-QAM (see Fig. \ref{figure:lambdamax}(b)). These results suggest that PS may impair the BPS performance for square-QAM constellations, but this behavior may change for other constellation geometries. 

\begin{figure*}
\centering
\includegraphics[scale = 1]{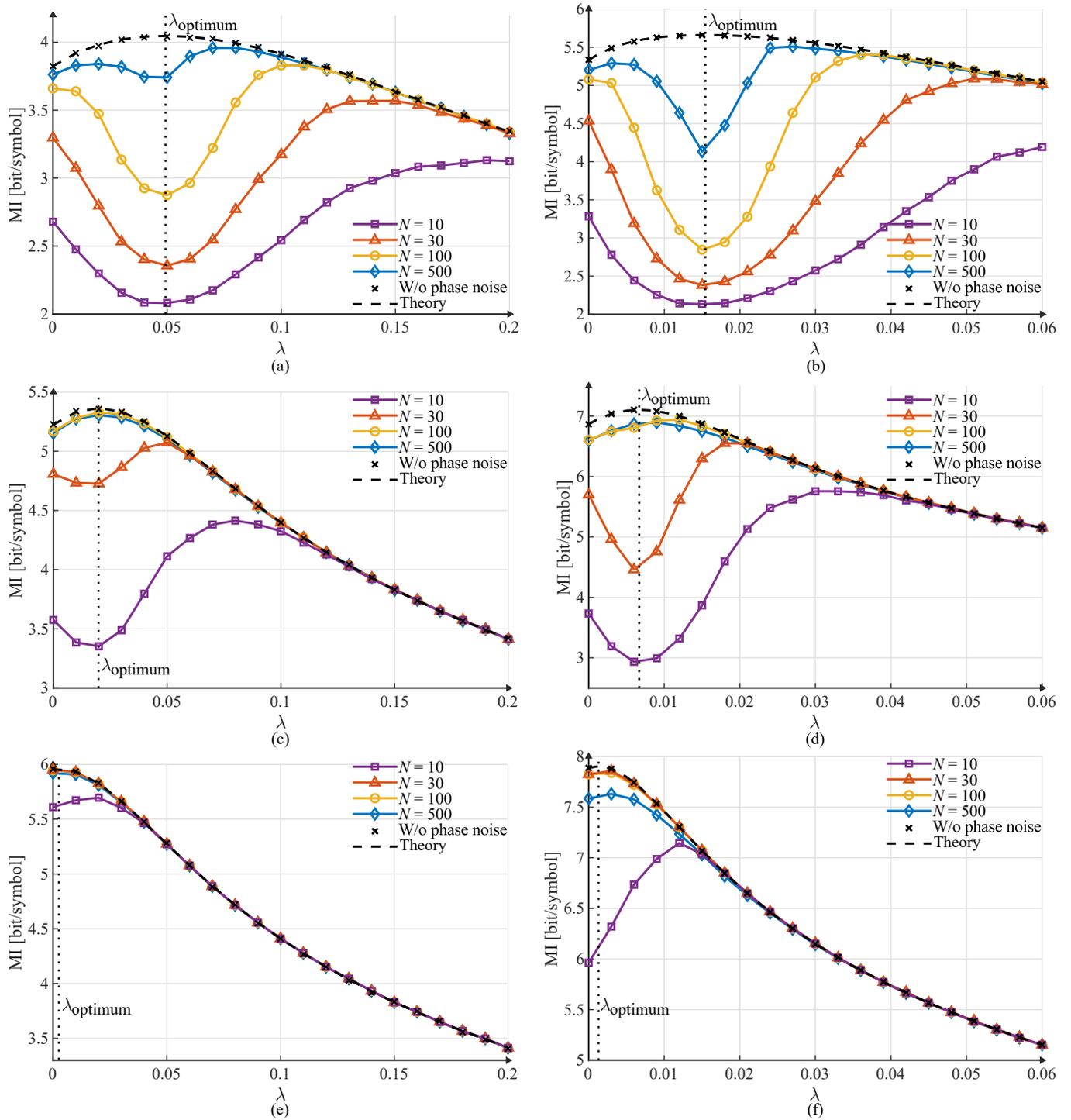}
\caption{MI evaluated with BPS assuming $N$ = 10, 30, 100, and 500, for PS-64-QAM at (a) SNR = 12 dB, (c) SNR = 17 dB, and (e) SNR = 22 dB; and for PS-256-QAM at (b) SNR = 17 dB, (d) SNR = 22 dB, and (f) SNR = 27 dB. The simulations include AWGN and Wiener phase noise corresponding to $\Delta\nu = 200$ kHz and a symbol rate of 50 GBd. Note that $\lambda=0$ corresponds to a uniform constellation.} \label{figure:MIversuslambda}
\end{figure*}

\begin{figure*}
\centering
\includegraphics[scale = 1]{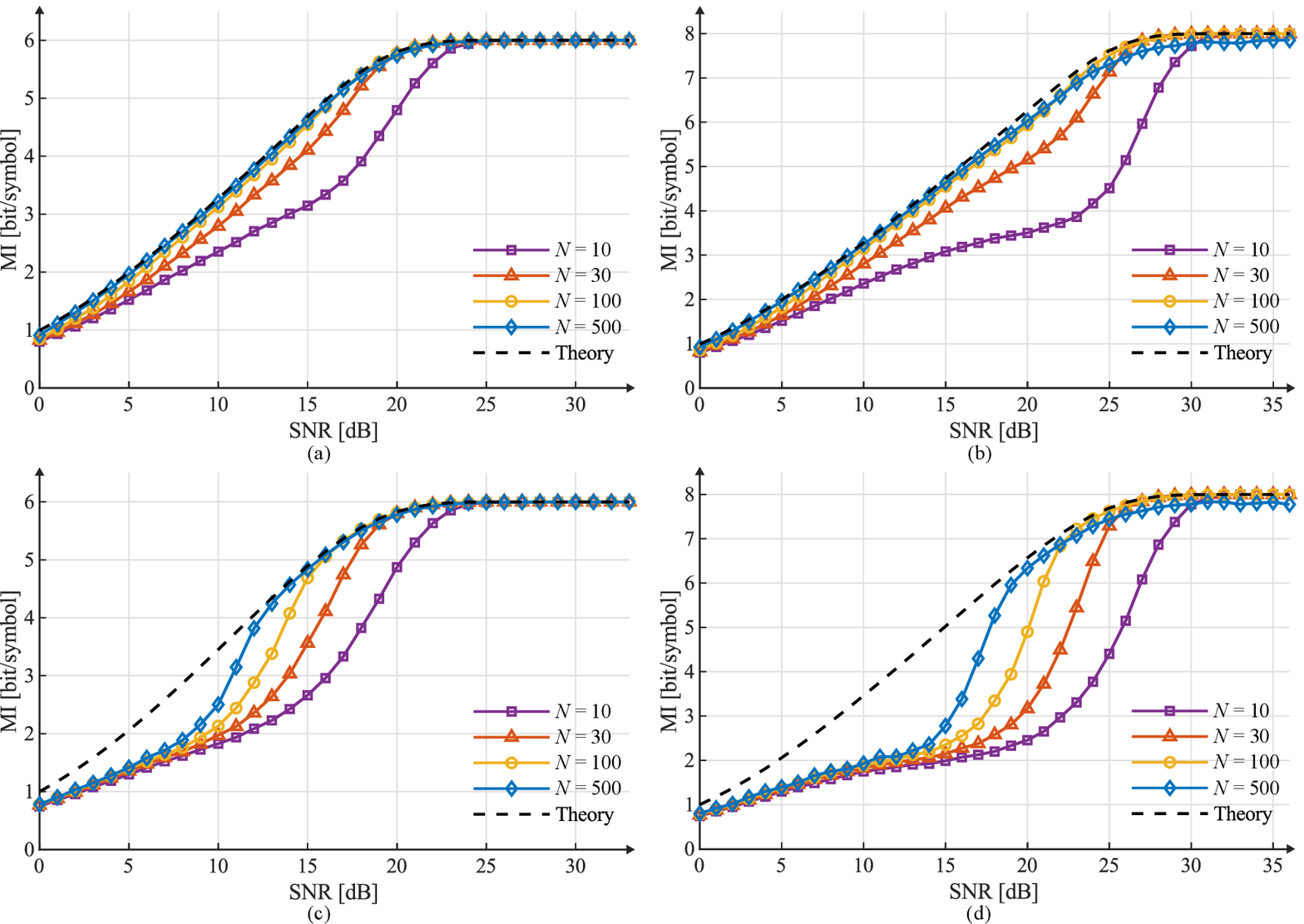}
\caption{MI as a function of SNR evaluated with BPS assuming $N$ = 10, 30, 100, and 500, for (a) 64-QAM, (b) 256-QAM, (c) PS-64-QAM, and (d) \mbox{PS-256-QAM}. The simulations include AWGN and Wiener phase noise corresponding to $\Delta\nu = 200$ kHz and a symbol rate of \mbox{50 GBd}. The shaping parameter $\lambda_{\textrm{optimum}}$ is used for all SNRs.} \label{figure:MISNR}
\end{figure*}

\subsection{MI performance}

In the previous section, we observed that the MSE of the estimated phase depends on the shaping parameter $\lambda$, and that the worst-case MSE is achieved in the vicinity of $\lambda_{\textrm{optimum}}$ for square-QAM constellations. In this section we assess the impact of this effect on the MI of a channel with Wiener phase noise. 
The simulations are performed with $2^{17}$ symbols. The BPS is implemented with 60 test phases. If left uncompensated, the occurrence of cycle slips in simulations with phase noise would disturb the estimation of the MI, which in this paper is based on the method used in \cite{Renner2017}. For this reason, we apply a supervised cycle slip compensation method that rotates every symbol at the output of BPS by multiples of $\pi/2$ to minimize the Euclidean distance to the corresponding transmitted symbol.

Fig. \ref{figure:MIversuslambda} evaluates the channel MI under the same noise levels and window sizes as in Fig. \ref{figure:mseunsupervised}. The dashed lines indicate the MI obtained numerically for an AWGN channel, and the crosses show simulation results used to validate the simulation setup. The simulated Wiener phase noise corresponds to  \mbox{$\Delta\nu$ = 200 kHz} and a symbol rate of 50 GBd.  It can be observed that errors in the phase estimation process cause a significant impact on the channel MI. For the lowest SNRs (Figs. \ref{figure:MIversuslambda}(a) and \ref{figure:MIversuslambda}(b)), in most cases the MI achieved without shaping ($\lambda$ = 0) is higher than that obtained with shaping. That is, instead of increasing the MI, the shaping causes the inverse effect of actually decreasing the channel MI. For example, this problem is clearly observed in Fig. \ref{figure:MIversuslambda}(a) for the 64-QAM format. The maximum theoretical MI for the channel without shaping (dashed line for $\lambda=0$) is reached for $N$ = 500, and almost reached for $N$ = 100. The MI for $N$ = 100 exhibits a sudden drop in $\lambda = \lambda_{\textrm{optimum}}$, making transmission uninteresting in this situation. Although the effect is milder for $N$ = 500, the MI reached at $\lambda = \lambda_{\textrm{optimum}}$ is still lower than that obtained in the uniform case ($\lambda=0$).  This situation is alleviated for intermediate SNR values (Figs. \ref{figure:MIversuslambda}(c) and \ref{figure:MIversuslambda}(d)), for which $N$ = 100 is sufficient to practically eliminate the impact of phase recovery on system performance. For high SNR values (Figs. \ref{figure:MIversuslambda}(e) and \ref{figure:MIversuslambda}(f)),  $N$ = 30 is sufficient to guarantee a suitable operation for both modulation formats, however, in this case, the shaping parameter is very low, and the constellation practically does not have shaping. It is interesting to note that,  for the PS-256-QAM format and \mbox{SNR = 27 dB} (Fig. \ref{figure:MIversuslambda}f), the curve for $N$ = 500 exhibits significant penalties because the noise rejection window is excessively long for the given balance of additive noise and phase noise.  

Possible SNR penalties due to phase recovery in shaped transmissions can be observed in Fig. \ref{figure:MISNR}, which shows the MI \emph{versus} SNR performance for uniform (top) and shaped (bottom) cases for both both 64-QAM (left) and 256-QAM (right) formats. For the uniform case, setting $N$ = 100 is enough to provide negligible implementation penalties for a wide range of SNRs. On the other hand, the shaped case exhibits steep drops, and even for long noise rejection windows the simulated curves detach from the theoretical ones at moderate SNR values, eliminating the expected SNR shaping gains.      

\begin{figure*}
\centering
\includegraphics[scale = 1]{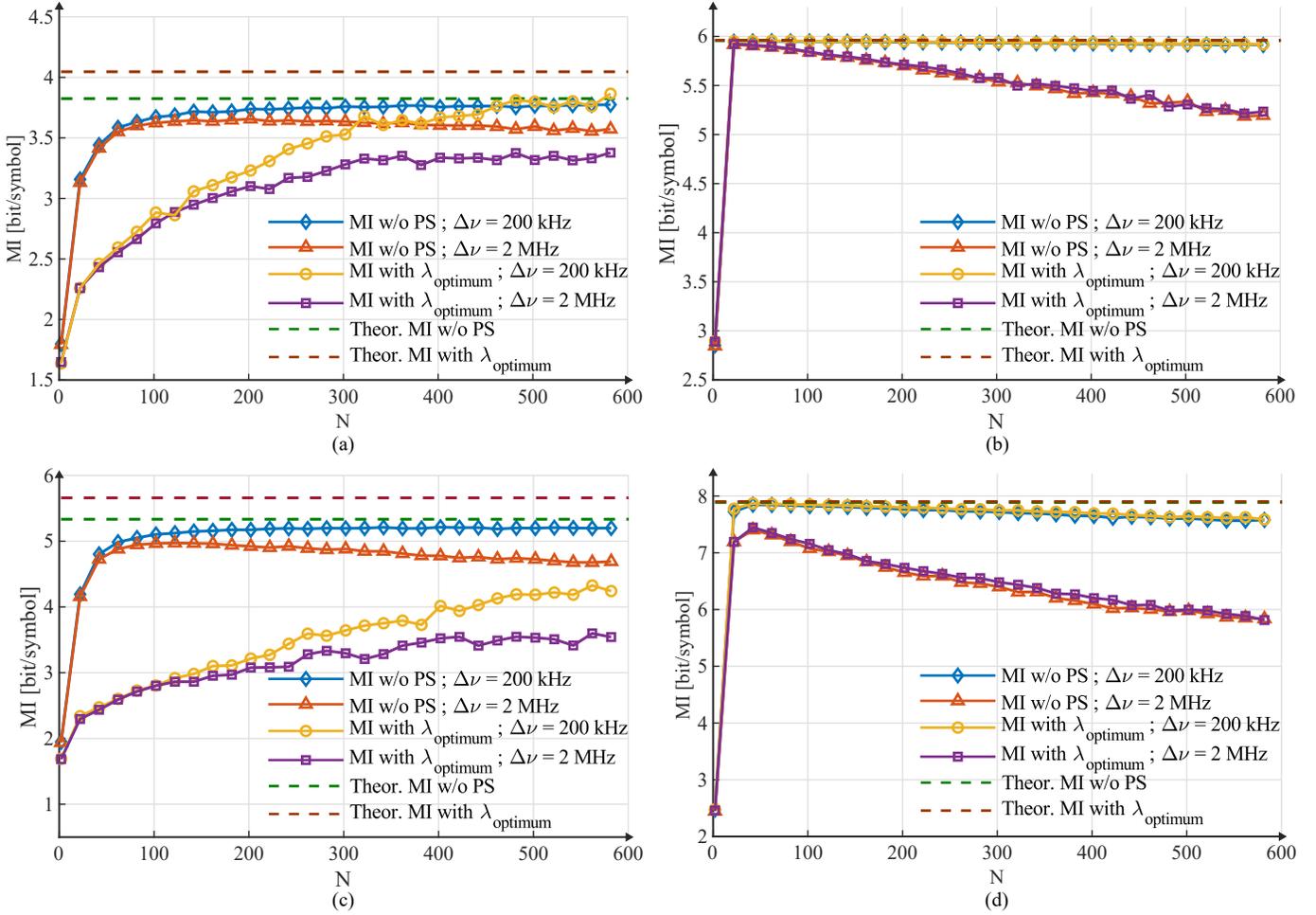}
\caption{MI as a function of $N$ for uniform and probabilistically shaped 64-QAM at (a) SNR = 12 dB and (b) SNR = 22 dB; and for uniform and probabilistically shaped 256-QAM at (c) SNR = 17 dB and (d) SNR = 27 dB. The simulations include AWGN and Wiener phase noise corresponding to a symbol rate of \mbox{50 GBd} and $\Delta\nu = 200$ kHz or $\Delta\nu = 2$ MHz.} \label{figure:wiener}
\end{figure*}

We also evaluated the MI as a function of the noise rejection window length for different values of SNR and laser linewidths. The results for the 64-QAM format are shown in Figs. \ref{figure:wiener}(a) (SNR = 12~dB) and \ref{figure:wiener}(b) (SNR = 22~dB). At SNR = 12~dB and uniform transmission the additive noise is dominant, and little dependence of the bit error rate on $N$ is observed, provided that the window is longer than approximately 200 symbols. Under these conditions, increasing the window size (e.g. up to 500) does not result in  system degradation, but increases the complexity and power consumption of the algorithm. The performance with probabilistic shaping is considerably poorer. Here, for a $\Delta\nu$ = 200 kHz, only \mbox{$N$ = 450} ensures a performance equivalent to the uniform case, and for $\Delta \nu$ = 2 MHz the performance of the uniform case is never reached. For  \mbox{SNR = 22 dB} the shaped and uniform cases coincide, as the shaping parameter is very low. A minimum window of approximately 20 symbols is sufficient to ensure adequate performance for both cases. However, using larger windows impairs the phase recovery process and consequently degrades the MI.  The performance for the 256-QAM format is shown in Figs. \ref{figure:wiener}(c) (SNR = 17~dB) and \ref{figure:wiener}(d) (SNR = 27~dB). For SNR = 17~dB without shaping, a filtering window of approximately 100 symbols is enough to achieve a relatively high MI. Again, PS strongly impairs the system performance. For both $\Delta \nu = 200$ kHz  and $\Delta \nu = 2$ MHz, the performance obtained by the uniform constellation is never achieved. For  SNR = 27 dB the shaped and uniform cases coincide, as the shaping parameter is very low. In this case, again, $N$ = 20 is enough to achieve the expected theoretical MI.

\section{Conclusion} \label{section:conclusion}

The interplay of PS and the BPS algorithm is investigated analytically and by simulation. We start by analyzing the performance of an SPS algorithm, which has the same architecture of BPS, except for the decision process, which is assumed perfect. We provide an analytical expression for the MSE of SPS, which exhibits a good agreement with simulations. The results demonstrate that PS affects the performance of SPS at short noise rejection windows, but this impact is easily mitigated at windows of moderate sizes. At large windows, the SPS MSE is independent on the modulation format and, thus, insensitive to PS. The BPS algorithm, however, reveals a strong dependence on PS, even for long noise rejection windows. Given the differences in  behavior of SPS and BPS, we infer that the decisions made inside the BPS algorithm are affected by shaping. For this reason, even long noise rejection windows may provide only modest gains to the algorithm performance. It is also observed that the worst shaping condition for the BPS algorithm is near to the capacity-maximizing operation point for square-QAM constellations. Finally, simulations of the MI of a channel with Wiener phase noise show that the PS impact on BPS can affect the overall system performance, specially at low SNRs. In this condition, the MI degradation caused by BPS can exceed potential capacity gains expected by PS. This effect can be eventually mitigated by extremely long noise rejection windows, which may increase complexity and require low linewidth lasers. These findings suggest the need for alternative phase recovery algorithms to be deployed in probabilistically-shaped transmissions. 

\section*{Acknowlegement} 

We would like to thank the editor and the anonymous reviewers for their essential contributions to improve the quality of the paper.

\appendix

\subsection{Derivation of the MSE for SPS}

A geometric analysis of the problem enables us to rewrite (\ref{equation:J}) as:
\begin{equation}
J(\theta_r) = \sum_{i=1}^N \left[\left(2|s_i|\textrm{sin}\left(\frac{\theta_n-\theta_r}{2}\right)+n_i^{(1)}\right)^2 + (n_i^{(2)})^2\right]  \\
\end{equation}
where $n_i^{(1)}$ is the noise component in the direction of the subtraction of $s_i$ and its rotated version $s_ie^{j(\theta_n-\theta_r)}$, and $n_i^{(2)}$ is the perpendicular component. Both  $n_i^{(1)}$ and $n_i^{(2)}$ are zero mean real Gaussian processes with variance $\sigma_n^2$ each. 
 
We find $\hat{\theta}_n$ by differentiating $J(\theta_r)$ with respect to $\theta_r$:
\begin{align}
& \frac{dJ(\theta_r)}{d\theta_r}   = \\ \nonumber
&\sum_{i=1}^N -2\left(2|s_i|\textrm{sin}\left(\frac{\theta_n-\theta_r}{2}\right)+n_i^{(1)}\right)|s_i|\textrm{cos}\left(\frac{\theta_n-\theta_r}{2}\right)
\end{align}
\indent Setting the derivative equal to zero, and supposing a small $\theta_n-\theta_r$, yields: 
\begin{align}
\sum_{i=1}^N \left(2|s_i|\textrm{sin}\left(\frac{\theta_n-\hat{\theta}_n}{2}\right)+n_i^{(1)}\right)|s_i|\approx 0\\
 \sum_{i=1}^N 2|s_i|^2\textrm{sin}\left(\frac{\theta_n-\hat{\theta}_n}{2}\right)+\sum_{i=1}^N (n_i^{(1)})|s_i| \approx 0\\
 \textrm{sin}\left(\frac{\theta_n-\hat{\theta}_n}{2}\right) \approx -\frac{1}{2}\frac{\sum_{i=1}^N (n_i^{(1)})|s_i|}{\sum_{i=1}^N |s_i|^2}
\end{align} 
\indent Approximating $\sin(x)\approx x$, gives:
\begin{align}
\frac{\theta_n-\hat{\theta}_n}{2} \approx -\frac{1}{2}\frac{\sum_{i=1}^N (n_i^{(1)})|s_i|}{\sum_{i=1}^N |s_i|^2}\\
\hat{\theta}_n \approx \theta_n + \frac{\sum_{i=1}^N (n_i^{(1)})|s_i|}{\sum_{i=1}^N |s_i|^2}
\end{align} 
\indent Finally, the mean squared error (MSE) in the estimation of $\theta_n$ can be given as:
\label{section:analytical}
\begin{equation}
 \displaystyle \textrm{MSE}_{\textrm{SPS}}(N) = E\{(\theta_n - \hat{\theta}_n )^2\} =   E\left \{\left[\frac{\sum_{i=1}^N (n_i^{(1)})|s_i|}{\sum_{i=1}^N |s_i|^2} \right]^2\right \}
\end{equation}
 
\bibliographystyle{IEEEtran}
\bibliography{references}

\end{document}